\documentclass[prb,twocolumn,superscriptaddress,longbibliography]{revtex4-1}

\usepackage{amsmath}
\usepackage{amssymb}
\usepackage{xspace}
\usepackage{graphicx}
\usepackage{grffile}
\usepackage{nicefrac}
\usepackage{color}

\begin{document}

\title{Theoretical design of highly correlated electron states in delafossite 
heterostructures}

\author{Frank Lechermann}
\affiliation{I. Institut f{\"u}r Theoretische Physik, Universit{\"a}t Hamburg, 
D-20355 Hamburg, Germany}
\author{Raphael Richter}
\affiliation{Institut f\"ur Technische Thermodynamik, Deutsches Zentrum f\"ur 
Luft- und Raumfahrt, D-70569 Stuttgart, Germany}

\begin{abstract}
Delafossites represent natural heterostructures which can host rather different
electronic characteristics in their constituting layers. The design of novel
heterostructure architectures highlighting the competition between such varying
layer properties is promising from the viewpoint of basic research as well
as for future technological applications. By means of the combination of density 
functional theory and dynamical mean-field theory, we here unveil the formation of 
highly correlated electron states in delafossite heterostructures build from 
metallic PdCrO$_2$ and insulating AgCrO$_2$. Due to the sophisticated coupling between 
layers of strong and of weak internal electron-electron interaction, 
correlation-induced semimetals at ambient temperature and doped Mott-insulators at lower 
temperature are predicted. The unique electronic structure of delafossite 
heterostructures opens a door to research on novel challenging quantum matter.
\end{abstract}


\maketitle

Delafossite oxides~\cite{sha71-1,*sha71-2,*sha71-3} $AB$O$_2$, where $A$ and $B$ 
are different metallic elements, have gained ample renewed interest, mainly from two
directions. First, the dominant part of these compounds is insulating,
and technologically promising due to their potential as transparent conducting 
oxides~\cite{kaw97} as well as their possibly multifunctional character, e.g. by 
showing magnetic/ferroelectric response~\cite{sek08,ter12} and high potential for 
photocatalytic applications~\cite{ouy08}. Second, the small subgroup of metallic 
delafossites usually displays an enormously high electric conductivity, belonging
to the most-conductive metals at room temperature~\cite{mac17,dao17}. 
General key signatures of delafossite crystal lattices are the triangular-coordinated planes 
and especially the dumbbell O$-A-$O bonds along the $c$-axis. The layered $R\bar{3}m$ 
architecture of alternating $A^+$- and $B^{3+}$O$_2$ planes gives rise to a 'natural 
heterostructure', i.e. the different planes may host a rather different electronic character. 
Tuning and designing the different aspects of delafossites by further technological 
heterostructuring thus may open a route to challenging new materials with exceptional 
properties.

Among these fascinating materials, the PdCrO$_2$ compound stands further out as
it unites high metallicity from the Pd$^+$($4d^9$) planes with Mott-insulating 
CrO$_2$ layers~\cite{tak09,noh14,lec18,sun18}. 
Replacing Pd by neighboring Ag formally leads to a closed $4d^{10}$ shell and 
nonsurprisingly, AgCrO$_2$ is indeed insulating with a technologically interesting band 
gap of 1.68\,eV~\cite{ouy08}. In both delafossites, the localized Cr spins order in
an antiferromagnetic 120$^\circ$ manner below N{\'e}el temperatures 37.5\,K (PdCrO$_2$),
21\,K (AgCrO$_2$). Moreover, the palladium delafossite shows an unconventional anomalous Hall
effect~\cite{tak10} at low temperature $T$, whereas the silver delafossite displays 
multiferroic properties~\cite{sek08}. While both 
individual compounds are of prominent interest of their own, we want to open a new path in 
delafossite research in this work. By using the oxide-heterostructure concept, the goal is to 
theoretically further engineer the unique layer-triggered competition between itinerant and 
insulating tendencies in these CrO$_2$-based delafossites. As we wil show, this manipulation 
enables the stabilization of even more intriguing electronic phases, and drives the 
challenge of the common metal-versus-insulator paradigm to another level.
\begin{figure}[t]
\includegraphics*[width=8.5cm]{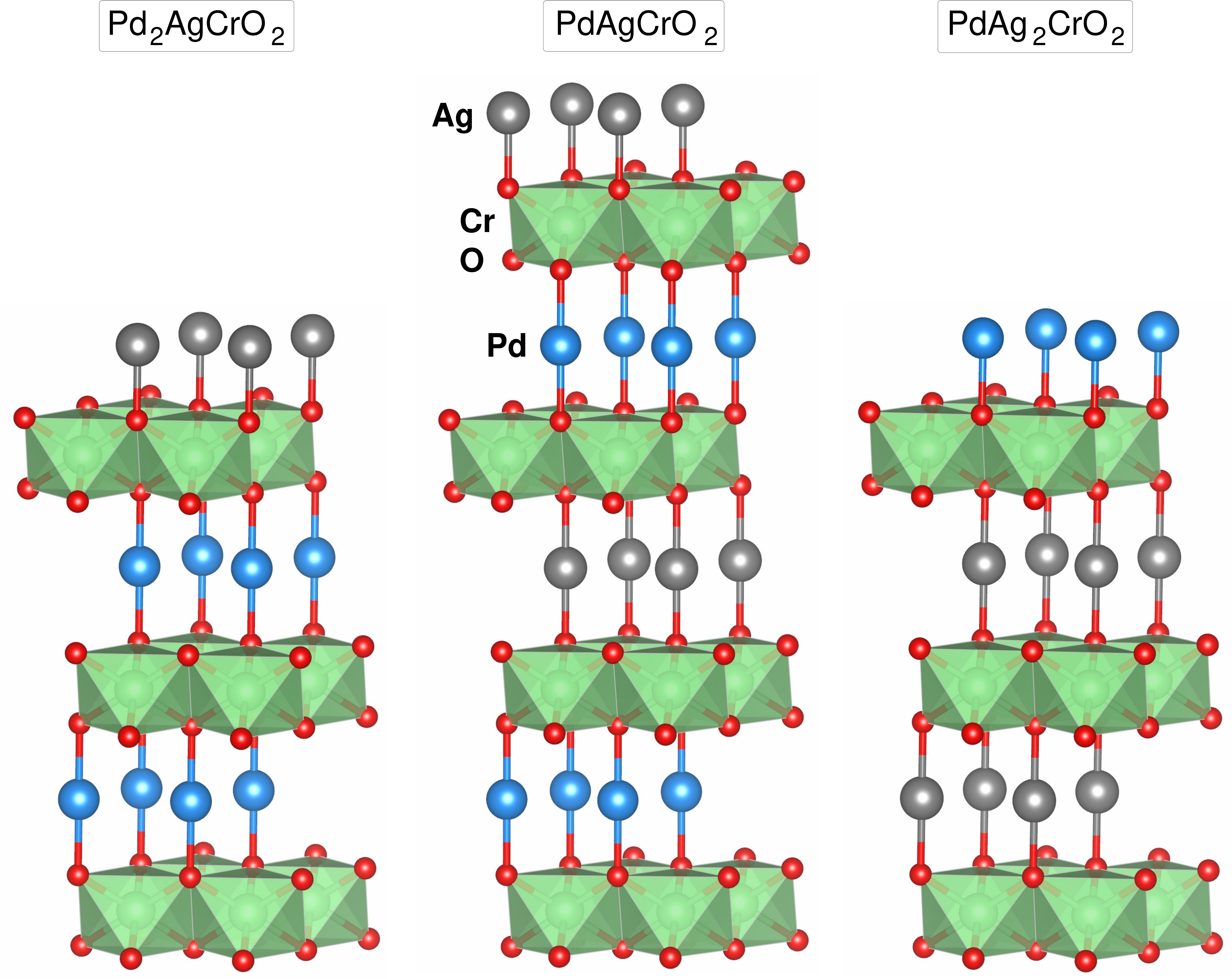}
\caption{(color online) Crystal structures of the designed delafossite heterostructures
with stacking along the $c$-axis. Pd$_2$AgCrO$_2$, PdAgCrO$_2$ and PdAg$_2$CrO$_2$ 
(from left to right). Pd: blue, Ag: grey, Cr: green and O: red.}
\label{fig:strucs}
\end{figure}
We study delafossite heterostructures Pd$_n$Ag$_m$CrO$_2$ with $n(m)$=1,2 planes of Pd(Ag)
in the primitive cell, and as in the conventional bulk compounds, separated by 
CrO$_2$ layers. Thus, three different artificial heterostructures are realized mediating
between the natural heterostructures PdCrO$_2$ and AgCrO$_2$, namely Pd$_2$AgCrO$_2$ (P2A), 
PdAgCrO$_2$ (PA) and PdAg$_2$CrO$_2$ (PA2) (see Fig.~\ref{fig:strucs}). In order to cope 
with the demanding problem involving the competition between itinerant, 
band-insulating as well as Mott-insulating layers, the elaborate
charge self-consistent combination~\cite{sav01,pou07,gri12} of density functional theory 
(DFT) and dynamical mean-field theory (DMFT) is put into practise to elucidate the  
correlated electronic structure, respectively. At room temperature, the three
designed heterostructures are predicted to be correlation-induced semimetals with 
singular electronic characteristics. Lowering the temperature leads to a doped-Mott
insulator scenario at low energy.

\section{Theoretical approach}
\begin{figure}[b]
\includegraphics*[width=8.5cm]{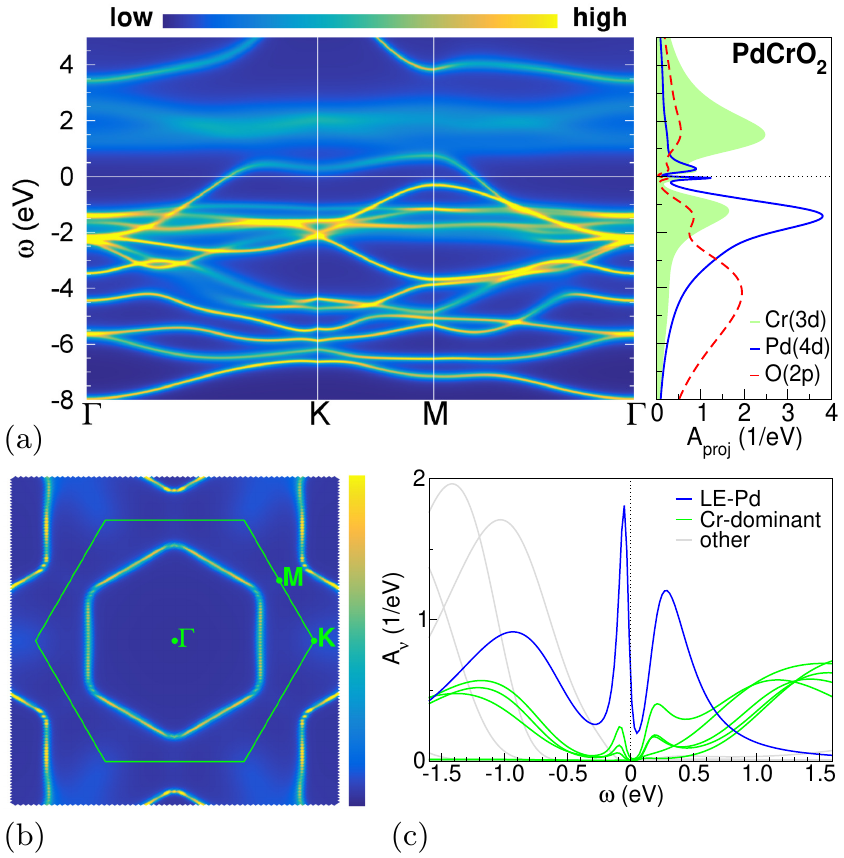}
\caption{(color online) Spectral data of PdCrO$_2$.
(a) Spectral function $A({\bf k},\omega)$ along high-symmetry lines in the $k_z=0$
plane of reciprocal space (left) and {\bf k}-integrated site- and orbital-projected 
spectral function (right).
(b) Fermi surface for $k_z=0$ within the first Brillouin zone (green hexagon) and
(c) {\bf k}-integrated Bloch contribution $A_{\nu}(\omega)$ (i.e. each individual
curve marks a different Bloch index $\nu$) with characterization of site dominance.
Note that the following figures use the same normalized color scale for the 
{\bf k}-resolved spectral function as (a) and (b).}
\label{fig:spectra-pd}
\end{figure}
\begin{figure}[b]
\includegraphics*[width=8.5cm]{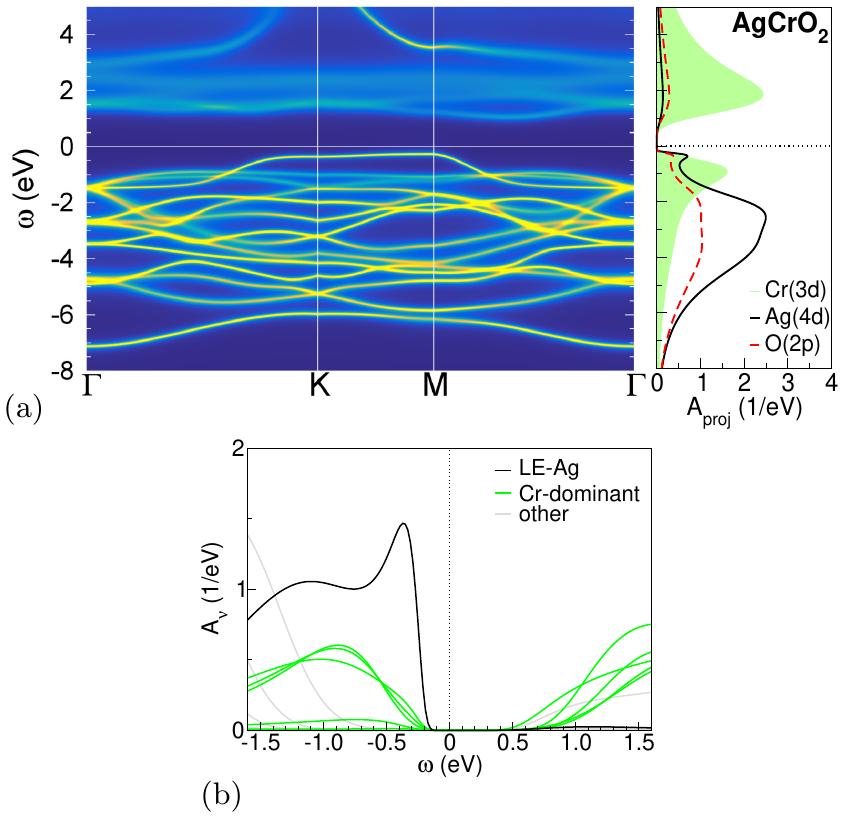}
\caption{(color online) Spectral data of AgCrO$_2$.
(a) Spectral function $A({\bf k},\omega)$ as in Fig~\ref{fig:spectra-pd}a.
(b) {\bf k}-integrated Bloch contribution $A_{\nu}(\omega)$ as in Fig~\ref{fig:spectra-pd}c.}
\label{fig:spectra-ag}
\end{figure}
A charge self-consistent combination of density functional theory and dynamical 
mean-field theory (DMFT) is employed~\cite{gri12}.
The mixed-basis pseudopotential method~\cite{lou79,mbpp_code}, based on norm-conserving 
pseudopotentials with a combined basis of localized functions and plane waves is used for
the DFT part with the generalized-gradient approximation in form of the PBE 
functional~\cite{per96}. Within the mixed basis, localized functions for Cr($3d$), 
Pd($4d$) and Ag($4d$) states as well as for O($2s$) and O($2p$) are utilized to reduce 
the plane-wave energy cutoff. The latter is chosen $E_{\rm cut}=20$\,Ryd for the bulk
structures and $E_{\rm cut}=16$\,Ryd for the heterostructures. A {\bf k}-point mesh of
13$\times$13$\times$13 partition is utilized for the bulk structures and of
11$\times$11$\times$3 partition for the heterostructures.

The correlated subspace for the DMFT part consists of the effective Cr($3d$) Wannier-like 
functions as obtained from the projected-local-orbital formalism~\cite{ama08,ani05}, 
using as projection functions the linear combinations of atomic $3d$ orbitals, 
diagonalizing the Cr($3d$) orbital-density matrix. A five-orbital
Slater-Kanamori Hubbard Hamiltonian governs the interacting electrons, parametrized by
a Hubbard $U$ and a Hund's exchange $J_{\rm H}$. 
The coupled single-site DMFT impurity problems in the given structures are solved by the 
continuous-time quantum Monte Carlo scheme~\cite{rub05,wer06} as implemented in the 
TRIQS package~\cite{par15,set16}. A double-counting correction of fully-localized-limit 
type~\cite{ani93} is applied. No Hubbard interactions are applied to Pd$(4d)$ and Ag$(4d)$.
Such interactions are already expected much smaller than on Cr$(3d)$ and then 
furthermore their effect onto nominal $d^9(d^{10})$ of Pd(Ag) should be weak in the 
metallic-layer architecture. Thus restricting the explicit DMFT-impurity treatment to 
the Cr sites is sound to study key correlation effects.

To obtain the spectral information, analytical continuation from Matsubara space via 
the maximum-entropy method as well as the Pad{\'e} method is performed. In the former
case, the {\bf k}-integrated Bloch Green's functions are continued, whereare in the 
latter case the local Cr self-energies are continued. If not otherwise stated, the 
maximum-entropy scheme is chosen to reveal {\bf k}-integrated spectral functions and the 
Pad{\'e} scheme is chosen to obtain {\bf k}-resolved spectral functions. Note moreover
that the {\bf k}-resolved spectral functions are computed on a dense mesh
along high-symmetry lines and in the $k_z=0$ plane of reciprocal space. For this purpose, 
the converged real-frequency local self-energies from the Pad{\'e} scheme, and the 
self-consistent projected-local-orbitals recomputed on the dense mesh, are utilized to 
obtain well-resolved Green's functions in {\bf k}-space.

The DFT+DMFT calculations are conducted by setting the system temperature to room 
temperature $T=290$\,K, if not otherwise stated. Paramagnetism is assumed in all the 
computations. Furthermore in the following, if not otherwise stated, all shown spectral 
data is obtained from charge self-consistent DFT+DMFT calculations.

\section{Results}
\subsection{PdCrO$_2$ and AgCrO$_2$\label{bulk}}
\begin{figure}[b]
\includegraphics*[width=8.5cm]{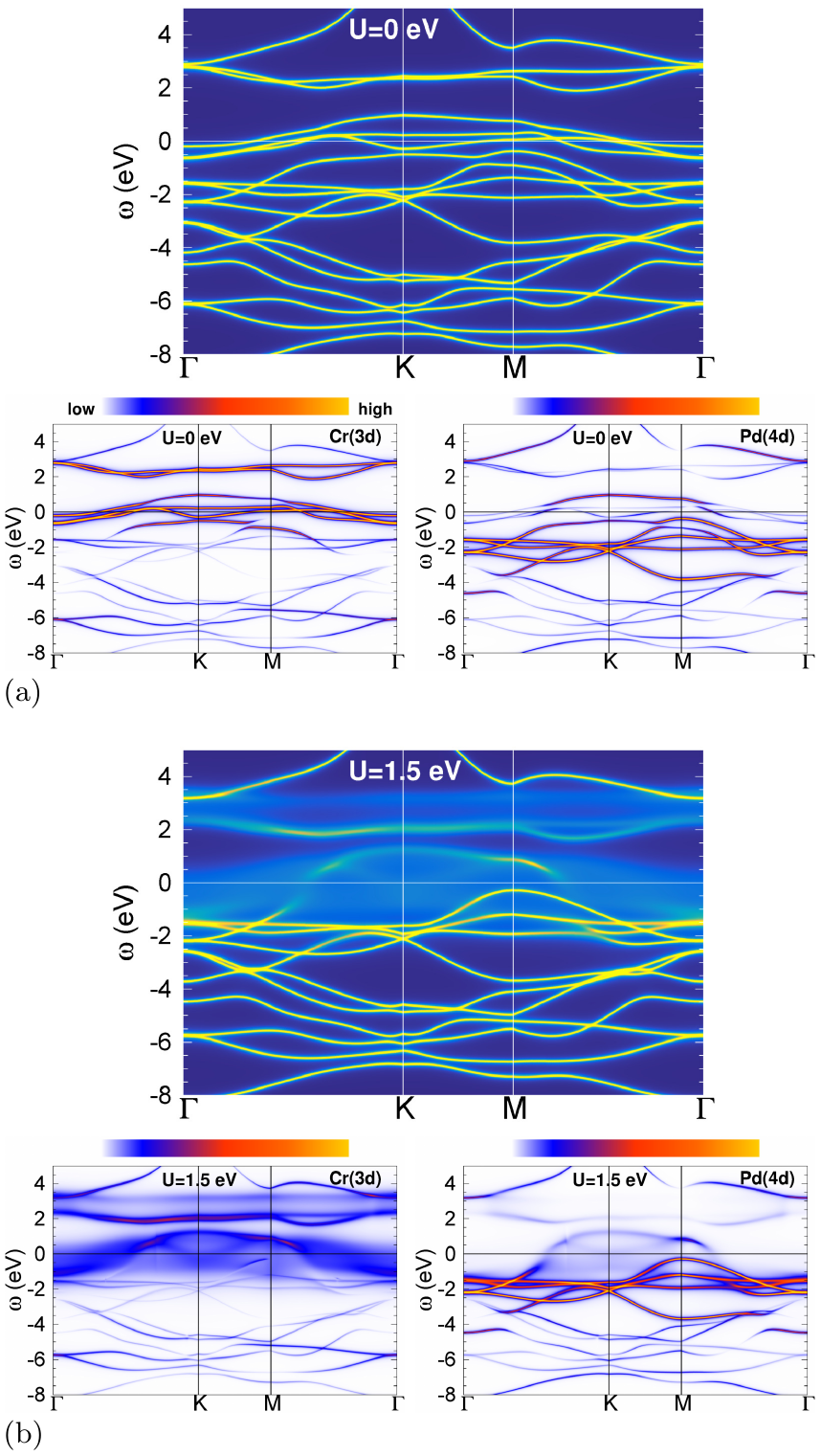}
\caption{(color online) {\bf k}-resolved spectral function (top) as well as Cr$(3d)$- and
Pd$(4d)$-resolved spectral-function weight (bottom) of PdCrO$_2$ for (a) 
$U=J_{\rm H}$=0 (i.e. DFT band structure) and for (b) $U=1.5$\,eV, $J_{\rm H}=0.35$\,eV.}
\label{fig:evol}
\end{figure}
To set the stage, we first briefly focus on the electronic structure of the natural 
bulk systems PdCrO$_2$ and AgCrO$_2$. Experimental lattice 
parameters~\cite{sha71-1,*sha71-2,*sha71-3,ouy08} $a=2.930$\,\AA\, and $c=18.087$\,\AA\, 
for PdCrO$_2$ as well as $a=2.985$\,\AA\, and $c=18.510$\,\AA\ for AgCrO$_2$ are used. 
The internal degree of freedom $z$, governing the oxygen distance to the $A^+$ plane is 
obtained from DFT structural optimization, reading $z=0.1101\,(0.1095)$ for 
PdCrO$_2$(AgCrO$_2$). Figures~\ref{fig:spectra-pd}, ~\ref{fig:spectra-ag} present a summary 
of the spectral characteristics. For PdCrO$_2$, local Coulomb interactions on the chromium 
site of $U=3$\,eV and $J_{\rm H}=0.7$\,eV prove adequate as discussed in 
Ref.~\onlinecite{lec18}. 
The system is metallic with a single conducting Pd-dominated (cPd) quasiparticle 
dispersion at the Fermi level $\varepsilon_{\rm F}$ and Mott-insulating 
CrO$_2$ layers (see Fig.~\ref{fig:spectra-pd}a). The cPd dispersion is dominantly formed by a 
linear combination of in-plane $d_{x^2-y^2}$ and $d_{xy}$, as well as by some $d_{z^2}$ 
contribution (see Ref.~\onlinecite{lec18} for more details).
\begin{figure*}[t]
\includegraphics*[width=17.8cm]{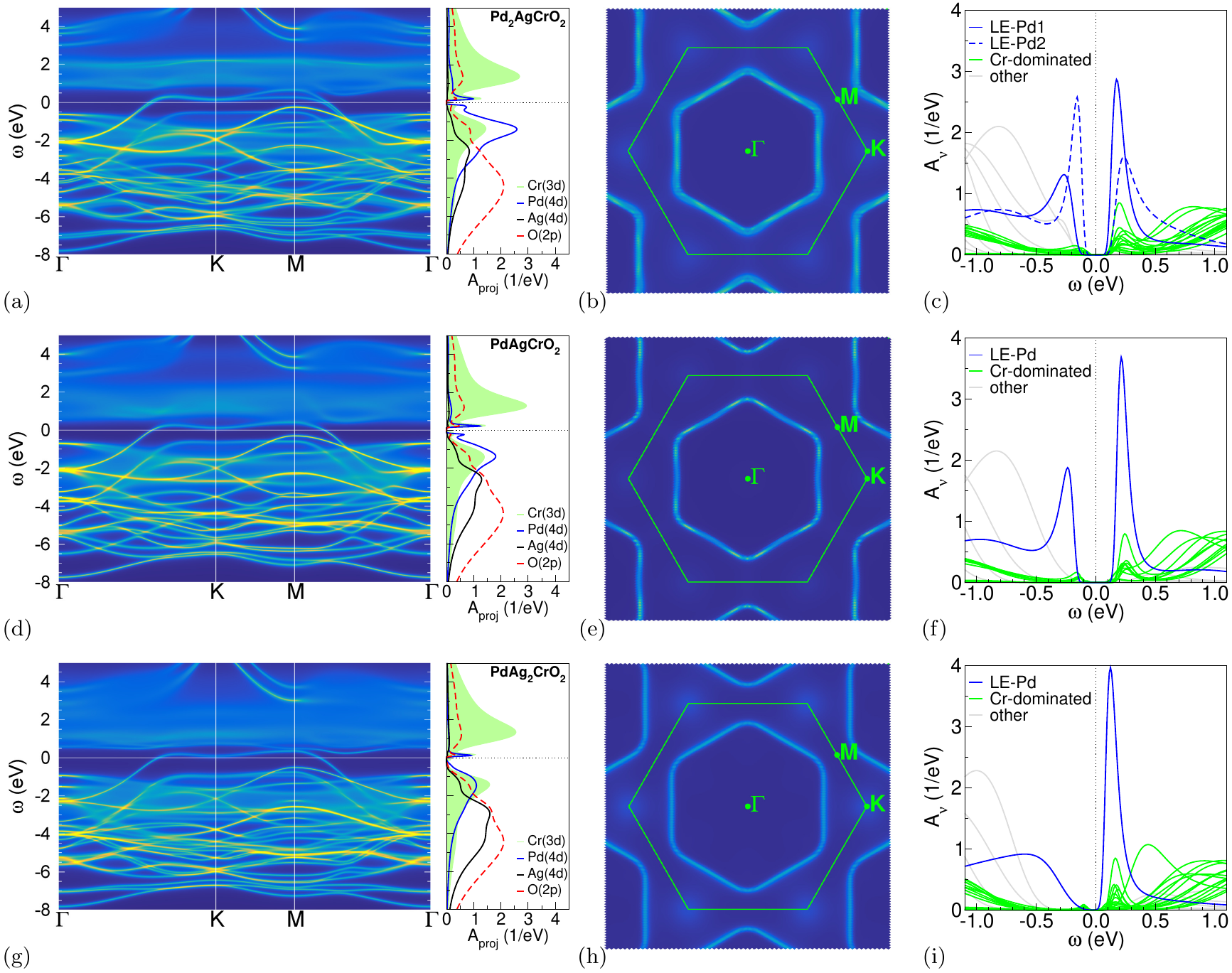}
\caption{(color online) Spectral data of the designed heterostructures: 
(a-c) Pd$_2$AgCrO$_2$, (d-f) PdAgCrO$_2$ and (g-i) PdAg$_2$CrO$_2$ at $T=290\,$K.
(a,d,g) Spectral function $A({\bf k},\omega)$ along high-symmetry lines in the $k_z=0$
plane of reciprocal space (left) and {\bf k}-integrated site- and orbital-projected 
spectral function (right). (b,e,h) Fermi surface for $k_z=0$ within the first 
Brillouin zone (green hexagon). (c,f,i) {\bf k}-integrated Bloch contribution 
$A_{\nu}(\omega)$ with characterization of dominance.
}
\label{fig:hetero}
\end{figure*}
Note that the {\bf k}-integrated projected spectral function $A_{\rm proj}(\omega)$ still 
displays sizable Cr-weight at low energy, associated with the cPd band. Thus intricate 
hybridization between the conducting and the Mott-insulating layers is at play~\cite{lec18}. 
The single-sheet Fermi surface, shown in Fig.~\ref{fig:spectra-pd}b, is of hole kind and of 
hexagonal shape, in line with photoemission~\cite{sob13,noh14}. It is furthermore 
instructive to visualize the spectral functions $A_{\nu}(\omega)$ with Bloch index $\nu$. 
The function $A_{\nu}(\omega)$ amounts to {\bf k}-integrated spectral weight of the 
correlated system per Bloch state, i.e. it provides the spectral information of
the correlated material as represented in Bloch Hilbert space. The total spectral function
then reads $A_{\rm tot}(\omega)=\sum_\nu A_{\nu}(\omega)$. Figure~\ref{fig:spectra-ag}c 
shows that dominantly, there is a single-$\nu$ low-energy Pd-dominated (LE-Pd) 
contribution at the Fermi level. Be still aware that the complete cPd spectral weight 
consists of the low-energy $\nu$-sum of all Bloch spectra.

In the case of AgCrO$_2$, the quasi closed-shell character of Ag reduces the screening within 
the CrO$_2$ subsystem and we therefore increased the Hubbard interaction on Cr to $U=4$\,eV. 
Figure~\ref{fig:spectra-ag}a,b exhibits that the CrO$_2$ layers remain Mott insulating, but the
previously conducting band becomes completely filled and thus the system as a whole 
insulating. Hence AgCrO$_2$ is a combined band-Mott insulator. The obtained charge gap of 
$\sim 1.8$\,eV is in good agreement with experiment. Note that the valence-band maximum is
of mixed Ag, O character, while the conduction-band minimum of dominant Cr upper-Hubbard
band character. Notably, whereas the Pd$(4d)$ states in PdCrO$_2$ align with the Cr lower
Hubbard band, the Ag$(4d)$ states align with O$(2p)$.

Let us comment on another important aspect, namely the evolution from the DFT 
electronic structure towards the fully-interacting one. When turning on interactions,
the electronic spectrum in CrO$_2$-based delafossites does not evolve from 
a non-interacting(-like) band structure to a Mott-insulating spectrum, as observed
for common Mott insulators~\cite{pav04}. For instance
in PdCrO$_2$, the metallic band structure evolves to another metallic spectrum, both
spectra with weakly-correlated-looking dispersions at low energy. This 
interaction evolution takes place in a highly nontrivial way, as depicted in 
Fig.~\ref{fig:evol}. In the DFT ($U=J_{\rm H}$=0) case, the three Cr-based $t_{2g}$ bands
cross the Fermi level and the later cPd band is still completely filled well below 
$\varepsilon_{\rm F}$, as visualized from the orbital-resolved spectral-function weights
at the bottom of Fig.~\ref{fig:evol}a. For $U=1.5$\,eV, $J_{\rm H}=0.35$\,eV, (i.e. half
the interaction strength of the true system) the CrO$_2$ layers are not yet Mott 
insulating, and the former Cr-based $t_{2g}$ bands are now strongly incoherent
even close to the Fermi level. Furthermore, the later cPd band also becomes very 
incoherent when entering the low-energy regime (cf. Fig.~\ref{fig:evol}b). These 
incoherent features may be observed from the spreading of the spectral weight 
(i.e. ``bluriness'') leading to a (near) disappearance of the original well-defined 
dispersions. Thus intriguingly, PdCrO$_2$ at hypothetical {\sl smaller} $U$ is a very 
incoherent metal (or ``bad metal''), apparently evading Fermi-liquid characteristics 
and seemingly in conflict with Luttinger's theorem. Coherence in the cPd band is reached 
at larger $U$ when the CrO$_2$ layers are Mott insulating, which again underlines the 
intricate coupling between the Pd and the CrO$_2$ layers. Substitutional doping of 
the Cr site by isovalent Mo could be a way to reach a lower effective $U$ value in
the system.

\subsection{Designed heterostructures}
For the heterostructures Pd$_n$Ag$_m$CrO$_2$, supercells of $1\times 1\times 3$ kind 
(12 basis atoms) for $n\ne m$ and of $1\times 1\times 2$ kind (6 basis atoms) for $n=m=1$ are 
constructed (see Fig.~\ref{fig:strucs}). The respective supercell lattice parameters are 
obtained from linear interpolation of the bulk parameters and the atomic positions are 
structurally optimized within DFT. The Hubbard $U$ on the Cr site is also chosen to result
from a linear interpolation of the respective bulk values, i.e. $J_{\rm H}=0.7$\,eV 
and $U$(P2A)$=3.33$\,eV, $U$(PA)$=3.50$\,eV, $U$(PA2)$=3.67$\,eV.

\subsubsection{At room temperature: $T=290$\,K}
In order to first address the phase stability, the electronic formation energy
\begin{eqnarray}
E_{\rm form}^{(nm)}=&&E_{\rm tot}(\mbox{Pd$_n$Ag$_m$CrO$_2$})\nonumber\\
&&-nE_{\rm tot}(\mbox{PdCrO$_2$})-mE_{\rm tot}(\mbox{AgCrO$_2$})\quad,
\end{eqnarray}
where $E_{\rm tot}$ are the total energies either in DFT or DFT+DMFT, may be considered.
Note that here, we aligned the plane-wave energy cutoff $E_{\rm cut}$ to obtain comparable 
total energies.
Within DFT, all designed heterostructures yield a negative $E_{\rm form}$, thus the considered 
phases are stable against decomposition into a mixture of the bulk phases. Within DFT+DMFT, 
only the P2A structure has a negative formation energy, but the absolute values of the order 
of $\sim 10$\,meV/atom are rather small troughout the series. Note that the given
$E_{\rm form}^{(nm)}$ should only be interpreted as a rough guide for phase stability,
since the latter depends furthermore on the temperature window, the entropy, the 
vibrational free energy and further nano-structuring effects, i.e. given through thin-film
architectures. Therefore in summary, we consider the given Pd$_n$Ag$_m$CrO$_2$ phases 
as indeed serious candidates for stable heterostructures in experiment.
\begin{figure}[b]
\includegraphics*[width=6.5cm]{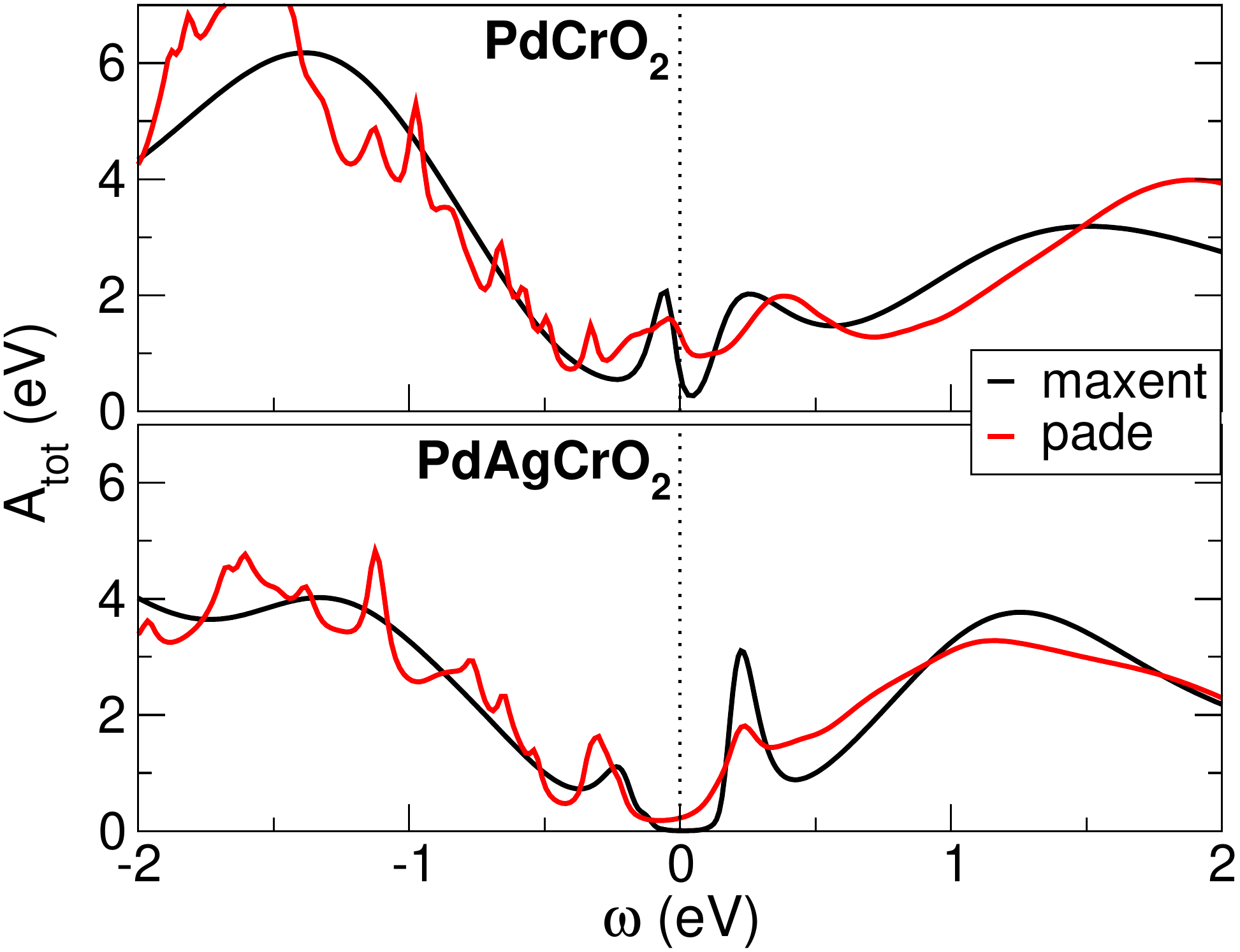}
\caption{(color online) Comparison between the total {\bf k}-integrated spectral functions
obtained from analytical continuation via maximum-entropy and via Pad{\'e} scheme for
PdCrO$_2$ (top) and PdAgCrO$_2$ (bottom). The displayed Pad{\'e} data results from
a straightforward {\bf k}-sum of $A({\bf k},\omega)$ on a uniform {\bf k}-mesh in
the respective Brillouin zone.}
\label{fig:comp}
\end{figure}
\begin{figure}[t]
\includegraphics*[width=8.5cm]{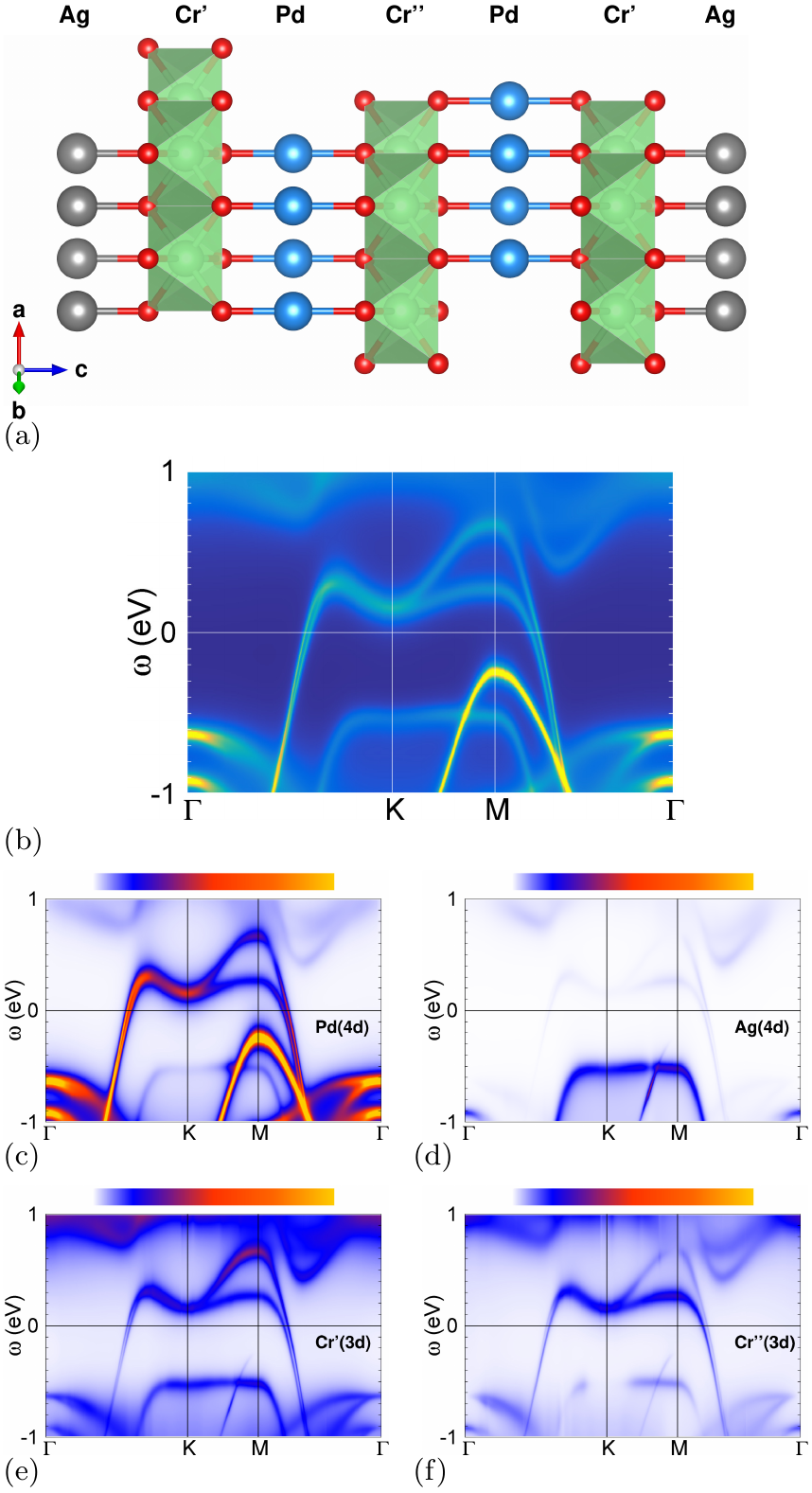}
\caption{(color online) Details on the Pd$_2$AgCrO$_2$ physics.
(a) Crystal structure with vertical $c$-axis, displaying the symmetry-inequivalent 
Cr' and Cr'' layers (color coding as Fig~\ref{fig:strucs}).
(b) Low-energy blow up of the ${\bf k}$-resolved spectral function along high-symmetry
lines. (c-f) Site- and orbital-resolved spectral-function weight: (c) Pd$(4d)$ 
(d) Ag$(4d)$, (e) Cr'$(3d)$ and (f) Cr''$(3d)$.}
\label{fig:p2a}
\end{figure}

In bulk PdCrO$_2$, the CrO$_2$ layers apparently do not severely harm the itinerancy of
the free Pd electron, the established Mott-insulating layers even tend to 
stabilize the weakly-correlated cPd dispersion (see last paragraph of 
section~\ref{bulk}). The present heterostructuring then formally poses 
a rather unique canonical problem: the single highly-itinerant electron, 
derived from the Pd layer, faces additional strong driving towards non-itinerant 
behavior imposed {\sl outside} its hosting layer via the blocking layers of coupled 
Ag-CrO$_2$ kind. Standard mechanisms of creating ``heaviness'' for such an electron 
are not easily applicable, yet simply ignoring those new boundary conditions is neither 
an option. How does the electron cope with the new situation at room temperature?

The spectral information for the three heterostructures is summarized
in Fig.~\ref{fig:hetero}. From the {\bf k}-resolved low-energy dispersions in 
Figs.~\ref{fig:hetero}a,d,g nothing spectacular seems to happen, there are still 
similar ``low-energy bands'' at $\varepsilon_{\rm F}$, reminiscent of the PdCrO$_2$ bulk
case. However surprisingly, the {\bf k}-integrated spectra (see also 
Figs.~\ref{fig:hetero}c,f,i) of all three
designed heterostructures show a pseudogap-like structure reaching zero spectral weight
at $\varepsilon_{\rm F}$. This gap feature is of size $\sim 0.2$\,eV for P2A and PA,
and about $\sim 0.1$\,eV for PA2. It is important to point out that 
this at first seemingly contradicting findings of {\bf k}-resolved vs. {\bf k}-integrated 
spectra are not an artifact of the analytical continuation from Matsubara to real 
frequencies, but well-reproducable both with maximum-entropy as well as with Pad{\'e} 
schemes. 
This is shown in Fig.~\ref{fig:comp}, depicting a comparison of the maximum-entropy vs. 
Pad{\'e}-derived spectrum for bulk PdCrO$_2$ and heterostructure PdAgCrO$_2$. 
While the straightforward {\bf k}-integrated Pad{\'e}-based spectrum is more 'wiggly', 
the key result at low energy is identical: whereas in PdCrO$_2$ a quasiparticle feature 
crosses $\varepsilon_{\rm F}$, PdAgCrO$_2$ shows a gap(-like) feature at $\omega=0$.

In fact, there is no contradiction between the {\bf k}-resovled and the {\bf k}-integrated
spectrum. The physical scenario is best coined by the term 
{\sl correlation-induced semimetal (CIS)}, where electronic correlations suppress the low-energy 
spectral weight in a non-trivial dispersion setting. In a correlated system, a dispersion
always carries spectral weight less than unity (contrary to e.g. a DFT picture), and in
the present case the spectral weight associated with the dispersion crossing 
$\varepsilon_{\rm F}$ is exceptionally small at room temperature. 
Figures~\ref{fig:hetero}c,f,i show that most of the expected low-energy spectral weight
piles up in sidebands at $\sim\pm 0.2$\,eV. Note that this specific low-energy
behavior is due to the unique coupling of originally weakly correlated Pd-based and 
strongly correlated Cr-based contributions in the (Pd,Ag)CrO$_2$ system. This coupling 
leads to low-energy dispersions of very weak spectral weight, resulting in neglible 
Fermi-level contribution upon ${\bf k}$ integration. 

Each Fermi-surface sheet, two in the case of P2A, still effectively encompasses one 
electron, respectively (cf. Figs.~\ref{fig:hetero}b,e,h). Only for the PA2 heterostructure 
the $\Gamma$-centered hole sheet, intriguingly, appears slightly enlarged, and we will come 
back to that observation in the next section.

A further aspect is noteworthy, and can be elaborated best from inspection of the P2A 
electronic structure with a subtle twofold dispersion crossing $\varepsilon_{\rm F}$
(cf. Fig.~\ref{fig:p2a}b). 
Figure~\ref{fig:p2a}a highlights the fact that there are two symmetry-inequivalent Cr 
layers: Cr'O$_2$ sandwiched by Ag and Pd layers as well as Cr''O$_2$ sandwiched by two Pd 
layers. The twofold dispersion crossing the Fermi level along $\Gamma$-K and M-$\Gamma$ is 
nonsuprisingly of dominant Pd$(4d)$ character as shown in Fig.~\ref{fig:p2a}c. The
low-energy Ag$(4d)$ weight accumulates in an occupied flat-band part along K-M with no
relevant contribution right at $\varepsilon_{\rm F}$ (see Fig.~\ref{fig:p2a}d). 
While the Cr' weight is balanced between the twofold dispersion crossing 
$\varepsilon_{\rm F}$, the Cr'' weight breaks this symmetry 
and attains somewhat larger weight on the FS sheet with slightly larger $k_{\rm F}$ 
(see Fig.~\ref{fig:p2a}e,f). Hence within the challenging correlation physics of the 
present delafossite heterostructures, there is furthermore an intricate hybridization 
component. 

\subsubsection{Below room temperature}
\begin{figure}[b]
\includegraphics*[width=8.5cm]{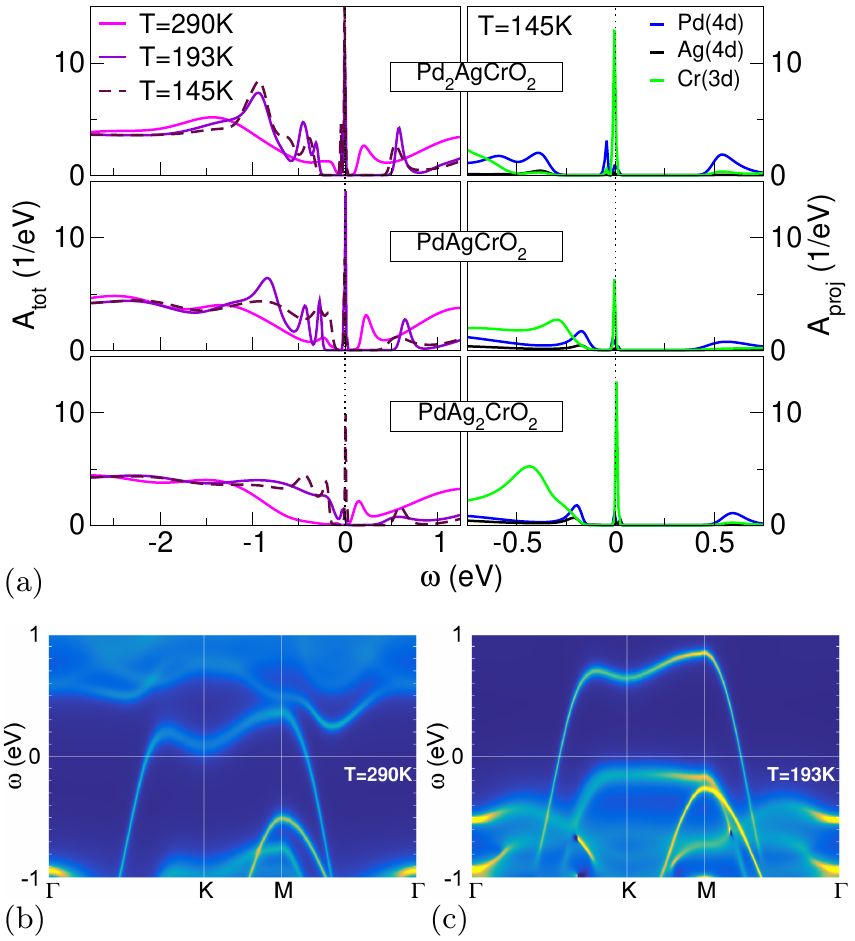}
\caption{(color online) Temperature-driven electronic transition in the delafossite 
heterostructures. (a) Evolution of the total spectral function with temperature
throughout the series. Left: $A_{\rm tot}(\omega)$ for P2A (top), PA (middle) and 
PA2 (bottom) at $T=290$\,K, $193$\,K and $145$\,K. 
Right: Element-specific projected spectral function at low energy for $T=145$\,K 
(same structure ordering as left part).
(b,c) Low-energy ${\bf k}$-resolved spectral function of PdAg$_2$CrO$_2$ for 
(b) $T=290$\,K and (c) $T=193$\,K.}
\label{fig:lowt}
\end{figure}

The CIS state at room temperature is expectedly the result of a very low coherence scale of 
an underlying effectively doped Mott-insulating state of the CrO$_2$ layers. The DFT+DMFT
calculations at lower temperature are numerically especially demanding for the multi-site
five-orbital correlated subspaces in the heterostructure design. But indeed, below
room temperature, a low-energy resonance emerges in the total spectral function of the 
delafossite heterostructures (see Fig.~\ref{fig:lowt}a).
Notably, this resonance is now of dominating Cr character, and carries sizable Pd weight
only for the P2A structure (see Fig.~\ref{fig:lowt}, right). Thus the given delafossite
heterostructures are rather sensitive to temperature and display a 
semimetal-to-doped-Mott-insulator transition around 190$\,$K. 

A minimal picture behind these numerical results may be as follows. Heterostructuring of
PdCrO$_2$ and AgCrO$_2$ leads to an effective doping of the CrO$_2$ layers, thus 
charge fluctuations between the Pd layers and the latter increase. Since the hopping
{\sl within} the Pd layers does not become heavily dressed in the Hubbard sense, the cPd 
dispersion is not strongly renormalized. In other words, there is no strong cPd 
band-narrowing effect, as seen in Fig.~\ref{fig:hetero}a,d,g. Instead, spectral-weight
transfer away from low energy and loss of coherence occurs for an only weakly
modified cPd dispersion. Still, the scattering with the doped CrO$_2$ layers is effective 
in marking a semimetallic scenario. At lower temperatures, the excitations in 
the doped-Mott layers eventually become coherent and a resonance appears
at low energy. 

Finally, let us remark on the fermiology of PA2. As obvious from Fig.~\ref{fig:lowt},
the PA2 structure develops a low-energy resonance at $T$ lower than the other two 
heterostructures. At room temperature, the PA2 structure is furthest away from the 
doped-Mott coherence state. Low-energy coherence of the Mott layers, in the sense 
previously discussed for Fig.~\ref{fig:evol}, is thus also not yet fully established at
that higher $T$. Therefore, the PA2 Fermi surface in Fig.~\ref{fig:hetero}h shows the
deviation from the expected sheet size. Only at lower temperature, when coherence is
reinforced, the expected sheet size for one electron is established. This is visualized
in the temperature comparison of the low-energy dispersions in Figs.~\ref{fig:lowt}b,c.

\section{Conclusions and Discussion}
Delafossite materials, and especially the metallic ones, are not only highly interesting 
in stoichiometric bulk form. In this paper we have shown that tailored heterostructuring 
of metallic PdCrO$_2$ and insulating AgCrO$_2$ gives rise to a plethora of
challening physics. The explicit aspect of electronic correlation, which is present but
to a large extent hidden within the bulk compounds, takes over in the heterostructure phases
Pd$_n$Ag$_m$CrO$_2$ and renders the demanding rivalry between metallicity and insulating
tendencies strikingly obvious. Correlation-induced semimetallic states at room temperature
and a transition to a doped Mott-insulating regime with low coherence scale at lower
temperature have been revealed. Furthermore, layer-selective hybridization may play a 
decisive role in order to select the stable many-body states at low energy. Eventual
competition with antiferromagnetic spin-ordering tendencies upon further lowering $T$ 
will even enlarge the scenario of competing many-body instabilities. But already without 
even considering possible magnetic ordering, there are many relevant degrees of freedom 
enabling a fascinating playground for very rich physics in designed delafossite
heterostructures. An interesting route for investigations could be given by 
uniaxial-pressure experiments~\cite{ric18,sun19,bar19}. Either to modify the Hubbard $U$ on 
Cr to reach the intricate non-Fermi-liquid regime at smaller $U$ (cf. Fig.~\ref{fig:evol}), 
or to effectively metallize the Mott layers by effective bandwidth enlargement.
Moreover, the construction of explicit topological states within a Mott-critical 
background may be possible from further tailored heterostructure design.

Still, the question concerning experimental realization of these and further delafossite
heterostructures arises. The representation of delafossites with standard growth techniques
has been proven challenging. However the recent growth of PdCoO$_2$ by molecular
beam epitaxy~\cite{bra19} is encouraging and should pave the road for a future realization
of delafossite heterostructures. In addition, single-layer deposition onto suitable 
terminated bulk delafossites may already prove helpful in creating exciting electron states
on a surface (see e.g. Ref.~\onlinecite{maz18} for work in that direction).

Let us conclude by the final remark that the key metal-delafossite physics of a crucial 
coupling between weakly itinerant layers and strongly correlated (Mott) layers connects
also to the recent finding of superconductivity in infinite-layer nickelate
upon hole doping~\cite{li19}, where similar scenarios might be at play at stoichiometry.

\begin{acknowledgments}
We thank A. P. Mackenzie and V. Sunko for helpful discussions. 
Financial support from the DFG LE-2446/4-1 project ``Design of strongly correlated
materials'' is acknowledged. Computations were performed at the 
JUWELS Cluster of the J\"{u}lich Supercomputing Centre (JSC) under project 
number hhh08. 
\end{acknowledgments}

\bibliography{bibextra}

\begin{thebibliography}{35}
\expandafter\ifx\csname natexlab\endcsname\relax\def\natexlab#1{#1}\fi
\expandafter\ifx\csname bibnamefont\endcsname\relax
  \def\bibnamefont#1{#1}\fi
\expandafter\ifx\csname bibfnamefont\endcsname\relax
  \def\bibfnamefont#1{#1}\fi
\expandafter\ifx\csname citenamefont\endcsname\relax
  \def\citenamefont#1{#1}\fi
\expandafter\ifx\csname url\endcsname\relax
  \def\url#1{\texttt{#1}}\fi
\expandafter\ifx\csname urlprefix\endcsname\relax\def\urlprefix{URL }\fi
\providecommand{\bibinfo}[2]{#2}
\providecommand{\eprint}[2][]{\url{#2}}

\bibitem[{\citenamefont{Shannon et~al.}(1971)\citenamefont{Shannon, Rogers, and
  Prewitt}}]{sha71-1}
\bibinfo{author}{\bibfnamefont{R.~D.} \bibnamefont{Shannon}},
  \bibinfo{author}{\bibfnamefont{D.~B.} \bibnamefont{Rogers}},
  \bibnamefont{and} \bibinfo{author}{\bibfnamefont{C.~T.}
  \bibnamefont{Prewitt}}, \bibinfo{journal}{Inorg. Chem.}
  \textbf{\bibinfo{volume}{10}}, \bibinfo{pages}{713} (\bibinfo{year}{1971}).

\bibitem[{\citenamefont{Prewitt et~al.}(1971)\citenamefont{Prewitt, Shannon,
  and Rogers}}]{sha71-2}
\bibinfo{author}{\bibfnamefont{C.~T.} \bibnamefont{Prewitt}},
  \bibinfo{author}{\bibfnamefont{R.~D.} \bibnamefont{Shannon}},
  \bibnamefont{and} \bibinfo{author}{\bibfnamefont{D.~B.}
  \bibnamefont{Rogers}}, \bibinfo{journal}{Inorg. Chem.}
  \textbf{\bibinfo{volume}{10}}, \bibinfo{pages}{719} (\bibinfo{year}{1971}).

\bibitem[{\citenamefont{Rogers et~al.}(1971)\citenamefont{Rogers, Shannon, and
  Prewitt}}]{sha71-3}
\bibinfo{author}{\bibfnamefont{D.~B.} \bibnamefont{Rogers}},
  \bibinfo{author}{\bibfnamefont{R.~D.} \bibnamefont{Shannon}},
  \bibnamefont{and} \bibinfo{author}{\bibfnamefont{C.~T.}
  \bibnamefont{Prewitt}}, \bibinfo{journal}{Inorg. Chem.}
  \textbf{\bibinfo{volume}{10}}, \bibinfo{pages}{723} (\bibinfo{year}{1971}).

\bibitem[{\citenamefont{Kawazoe et~al.}(1997)\citenamefont{Kawazoe, Yasukawa,
  Hyodo, Kurita, Yanagi, and Hosono}}]{kaw97}
\bibinfo{author}{\bibfnamefont{H.}~\bibnamefont{Kawazoe}},
  \bibinfo{author}{\bibfnamefont{M.}~\bibnamefont{Yasukawa}},
  \bibinfo{author}{\bibfnamefont{H.}~\bibnamefont{Hyodo}},
  \bibinfo{author}{\bibfnamefont{M.}~\bibnamefont{Kurita}},
  \bibinfo{author}{\bibfnamefont{H.}~\bibnamefont{Yanagi}}, \bibnamefont{and}
  \bibinfo{author}{\bibfnamefont{H.}~\bibnamefont{Hosono}},
  \bibinfo{journal}{Nature} \textbf{\bibinfo{volume}{389}},
  \bibinfo{pages}{939} (\bibinfo{year}{1997}).

\bibitem[{\citenamefont{Seki et~al.}(2008)\citenamefont{Seki, Onose, and
  Tokura}}]{sek08}
\bibinfo{author}{\bibfnamefont{S.}~\bibnamefont{Seki}},
  \bibinfo{author}{\bibfnamefont{Y.}~\bibnamefont{Onose}}, \bibnamefont{and}
  \bibinfo{author}{\bibfnamefont{Y.}~\bibnamefont{Tokura}},
  \bibinfo{journal}{Phys. Rev. Lett.} \textbf{\bibinfo{volume}{101}},
  \bibinfo{pages}{067204} (\bibinfo{year}{2008}).

\bibitem[{\citenamefont{Terada et~al.}(2012)\citenamefont{Terada, Khalyavin,
  Manuel, Tsujimoto, Knight, Radaelli, Suzuki, and Kitazawa}}]{ter12}
\bibinfo{author}{\bibfnamefont{N.}~\bibnamefont{Terada}},
  \bibinfo{author}{\bibfnamefont{D.~D.} \bibnamefont{Khalyavin}},
  \bibinfo{author}{\bibfnamefont{P.}~\bibnamefont{Manuel}},
  \bibinfo{author}{\bibfnamefont{Y.}~\bibnamefont{Tsujimoto}},
  \bibinfo{author}{\bibfnamefont{K.}~\bibnamefont{Knight}},
  \bibinfo{author}{\bibfnamefont{P.~G.} \bibnamefont{Radaelli}},
  \bibinfo{author}{\bibfnamefont{H.~S.} \bibnamefont{Suzuki}},
  \bibnamefont{and} \bibinfo{author}{\bibfnamefont{H.}~\bibnamefont{Kitazawa}},
  \bibinfo{journal}{Phys. Rev. Lett.} \textbf{\bibinfo{volume}{109}},
  \bibinfo{pages}{097203} (\bibinfo{year}{2012}).

\bibitem[{\citenamefont{Ouyang et~al.}(2008)\citenamefont{Ouyang, Li, Ouyang,
  Yu, Ye, and Zou}}]{ouy08}
\bibinfo{author}{\bibfnamefont{S.}~\bibnamefont{Ouyang}},
  \bibinfo{author}{\bibfnamefont{Z.}~\bibnamefont{Li}},
  \bibinfo{author}{\bibfnamefont{Z.}~\bibnamefont{Ouyang}},
  \bibinfo{author}{\bibfnamefont{T.}~\bibnamefont{Yu}},
  \bibinfo{author}{\bibfnamefont{J.}~\bibnamefont{Ye}}, \bibnamefont{and}
  \bibinfo{author}{\bibfnamefont{Z.}~\bibnamefont{Zou}}, \bibinfo{journal}{J.
  Phys. Chem. C} \textbf{\bibinfo{volume}{112}}, \bibinfo{pages}{3134}
  (\bibinfo{year}{2008}).

\bibitem[{\citenamefont{Mackenzie}(2017)}]{mac17}
\bibinfo{author}{\bibfnamefont{A.~P.} \bibnamefont{Mackenzie}},
  \bibinfo{journal}{Rep. Prog. Phys.} \textbf{\bibinfo{volume}{80}},
  \bibinfo{pages}{032501} (\bibinfo{year}{2017}).

\bibitem[{\citenamefont{Daou et~al.}(2017)\citenamefont{Daou, Fr{\'e}sard,
  Eyert, H{\'e}bert, and Maignan}}]{dao17}
\bibinfo{author}{\bibfnamefont{R.}~\bibnamefont{Daou}},
  \bibinfo{author}{\bibfnamefont{R.}~\bibnamefont{Fr{\'e}sard}},
  \bibinfo{author}{\bibfnamefont{V.}~\bibnamefont{Eyert}},
  \bibinfo{author}{\bibfnamefont{S.}~\bibnamefont{H{\'e}bert}},
  \bibnamefont{and} \bibinfo{author}{\bibfnamefont{A.}~\bibnamefont{Maignan}},
  \bibinfo{journal}{Science and Technology of Advanced Materials}
  \textbf{\bibinfo{volume}{18}}, \bibinfo{pages}{919} (\bibinfo{year}{2017}).

\bibitem[{\citenamefont{Takatsu et~al.}(2009)\citenamefont{Takatsu, Yoshizawa,
  Yonezawa, and Maeno}}]{tak09}
\bibinfo{author}{\bibfnamefont{H.}~\bibnamefont{Takatsu}},
  \bibinfo{author}{\bibfnamefont{H.}~\bibnamefont{Yoshizawa}},
  \bibinfo{author}{\bibfnamefont{S.}~\bibnamefont{Yonezawa}}, \bibnamefont{and}
  \bibinfo{author}{\bibfnamefont{Y.}~\bibnamefont{Maeno}},
  \bibinfo{journal}{Phys. Rev. B} \textbf{\bibinfo{volume}{79}},
  \bibinfo{pages}{104424} (\bibinfo{year}{2009}).

\bibitem[{\citenamefont{Noh et~al.}(2014)\citenamefont{Noh, Jeong, Chang,
  Jeong, Moon, Cho, Ok, Kim, Kim, Min et~al.}}]{noh14}
\bibinfo{author}{\bibfnamefont{H.-J.} \bibnamefont{Noh}},
  \bibinfo{author}{\bibfnamefont{J.}~\bibnamefont{Jeong}},
  \bibinfo{author}{\bibfnamefont{B.}~\bibnamefont{Chang}},
  \bibinfo{author}{\bibfnamefont{D.}~\bibnamefont{Jeong}},
  \bibinfo{author}{\bibfnamefont{H.~S.} \bibnamefont{Moon}},
  \bibinfo{author}{\bibfnamefont{E.-J.} \bibnamefont{Cho}},
  \bibinfo{author}{\bibfnamefont{J.~M.} \bibnamefont{Ok}},
  \bibinfo{author}{\bibfnamefont{J.~S.} \bibnamefont{Kim}},
  \bibinfo{author}{\bibfnamefont{K.}~\bibnamefont{Kim}},
  \bibinfo{author}{\bibfnamefont{B.~I.} \bibnamefont{Min}},
  \bibnamefont{et~al.}, \bibinfo{journal}{Sci. Rep.}
  \textbf{\bibinfo{volume}{4}}, \bibinfo{pages}{3680} (\bibinfo{year}{2014}).

\bibitem[{\citenamefont{Lechermann}(2018)}]{lec18}
\bibinfo{author}{\bibfnamefont{F.}~\bibnamefont{Lechermann}},
  \bibinfo{journal}{Phys. Rev. Materials} \textbf{\bibinfo{volume}{2}},
  \bibinfo{pages}{085004} (\bibinfo{year}{2018}).

\bibitem[{\citenamefont{Sunko et~al.}(2020)\citenamefont{Sunko, Mazzola,
  Kitamura, Khim, Kushwaha, Clark, Watson, Markovic, Biswas, Pourovskii
  et~al.}}]{sun18}
\bibinfo{author}{\bibfnamefont{V.}~\bibnamefont{Sunko}},
  \bibinfo{author}{\bibfnamefont{F.}~\bibnamefont{Mazzola}},
  \bibinfo{author}{\bibfnamefont{S.}~\bibnamefont{Kitamura}},
  \bibinfo{author}{\bibfnamefont{S.}~\bibnamefont{Khim}},
  \bibinfo{author}{\bibfnamefont{P.}~\bibnamefont{Kushwaha}},
  \bibinfo{author}{\bibfnamefont{O.~J.} \bibnamefont{Clark}},
  \bibinfo{author}{\bibfnamefont{M.}~\bibnamefont{Watson}},
  \bibinfo{author}{\bibfnamefont{I.}~\bibnamefont{Markovic}},
  \bibinfo{author}{\bibfnamefont{D.}~\bibnamefont{Biswas}},
  \bibinfo{author}{\bibfnamefont{L.}~\bibnamefont{Pourovskii}},
  \bibnamefont{et~al.}, \bibinfo{journal}{Sci. Adv.}
  \textbf{\bibinfo{volume}{6}}, \bibinfo{pages}{eaaz0611}
  (\bibinfo{year}{2020}).

\bibitem[{\citenamefont{Takatsu et~al.}(2010)\citenamefont{Takatsu, Yonezawa,
  Michioka, Yoshimura, and Maeno}}]{tak10}
\bibinfo{author}{\bibfnamefont{H.}~\bibnamefont{Takatsu}},
  \bibinfo{author}{\bibfnamefont{S.}~\bibnamefont{Yonezawa}},
  \bibinfo{author}{\bibfnamefont{C.}~\bibnamefont{Michioka}},
  \bibinfo{author}{\bibfnamefont{K.}~\bibnamefont{Yoshimura}},
  \bibnamefont{and} \bibinfo{author}{\bibfnamefont{Y.}~\bibnamefont{Maeno}},
  \bibinfo{journal}{J. Phys. Conf. Ser.} \textbf{\bibinfo{volume}{200}},
  \bibinfo{pages}{012198} (\bibinfo{year}{2010}).

\bibitem[{\citenamefont{Savrasov et~al.}(2001)\citenamefont{Savrasov, Kotliar,
  and Abrahams}}]{sav01}
\bibinfo{author}{\bibfnamefont{S.~Y.} \bibnamefont{Savrasov}},
  \bibinfo{author}{\bibfnamefont{G.}~\bibnamefont{Kotliar}}, \bibnamefont{and}
  \bibinfo{author}{\bibfnamefont{E.}~\bibnamefont{Abrahams}},
  \bibinfo{journal}{Nature} \textbf{\bibinfo{volume}{410}},
  \bibinfo{pages}{793} (\bibinfo{year}{2001}).

\bibitem[{\citenamefont{Pourovskii et~al.}(2007)\citenamefont{Pourovskii,
  Amadon, Biermann, and Georges}}]{pou07}
\bibinfo{author}{\bibfnamefont{L.~V.} \bibnamefont{Pourovskii}},
  \bibinfo{author}{\bibfnamefont{B.}~\bibnamefont{Amadon}},
  \bibinfo{author}{\bibfnamefont{S.}~\bibnamefont{Biermann}}, \bibnamefont{and}
  \bibinfo{author}{\bibfnamefont{A.}~\bibnamefont{Georges}},
  \bibinfo{journal}{Phys. Rev. B} \textbf{\bibinfo{volume}{76}},
  \bibinfo{pages}{235101} (\bibinfo{year}{2007}).

\bibitem[{\citenamefont{Grieger et~al.}(2012)\citenamefont{Grieger, Piefke,
  Peil, and Lechermann}}]{gri12}
\bibinfo{author}{\bibfnamefont{D.}~\bibnamefont{Grieger}},
  \bibinfo{author}{\bibfnamefont{C.}~\bibnamefont{Piefke}},
  \bibinfo{author}{\bibfnamefont{O.~E.} \bibnamefont{Peil}}, \bibnamefont{and}
  \bibinfo{author}{\bibfnamefont{F.}~\bibnamefont{Lechermann}},
  \bibinfo{journal}{Phys. Rev. B} \textbf{\bibinfo{volume}{86}},
  \bibinfo{pages}{155121} (\bibinfo{year}{2012}).

\bibitem[{\citenamefont{Louie et~al.}(1979)\citenamefont{Louie, Ho, and
  Cohen}}]{lou79}
\bibinfo{author}{\bibfnamefont{S.~G.} \bibnamefont{Louie}},
  \bibinfo{author}{\bibfnamefont{K.~M.} \bibnamefont{Ho}}, \bibnamefont{and}
  \bibinfo{author}{\bibfnamefont{M.~L.} \bibnamefont{Cohen}},
  \bibinfo{journal}{Phys. Rev. B} \textbf{\bibinfo{volume}{19}},
  \bibinfo{pages}{1774} (\bibinfo{year}{1979}).

\bibitem[{\citenamefont{Meyer et~al.}(1998)\citenamefont{Meyer, Els\"{a}sser,
  Lechermann, and F\"{a}hnle}}]{mbpp_code}
\bibinfo{author}{\bibfnamefont{B.}~\bibnamefont{Meyer}},
  \bibinfo{author}{\bibfnamefont{C.}~\bibnamefont{Els\"{a}sser}},
  \bibinfo{author}{\bibfnamefont{F.}~\bibnamefont{Lechermann}},
  \bibnamefont{and}
  \bibinfo{author}{\bibfnamefont{M.}~\bibnamefont{F\"{a}hnle}},
  \emph{\bibinfo{title}{FORTRAN 90 Program for Mixed-Basis-Pseudopotential
  Calculations for Crystals}}, \bibinfo{organization}{Max-Planck-Institut
  f\"{u}r Metallforschung, Stuttgart} (\bibinfo{year}{1998}).

\bibitem[{\citenamefont{Perdew et~al.}(1996)\citenamefont{Perdew, Burke, and
  Ernzerhof}}]{per96}
\bibinfo{author}{\bibfnamefont{J.~P.} \bibnamefont{Perdew}},
  \bibinfo{author}{\bibfnamefont{K.}~\bibnamefont{Burke}}, \bibnamefont{and}
  \bibinfo{author}{\bibfnamefont{M.}~\bibnamefont{Ernzerhof}},
  \bibinfo{journal}{Phys. Rev. Lett.} \textbf{\bibinfo{volume}{77}},
  \bibinfo{pages}{3865} (\bibinfo{year}{1996}).

\bibitem[{\citenamefont{Amadon et~al.}(2008)\citenamefont{Amadon, Lechermann,
  Georges, Jollet, Wehling, and Lichtenstein}}]{ama08}
\bibinfo{author}{\bibfnamefont{B.}~\bibnamefont{Amadon}},
  \bibinfo{author}{\bibfnamefont{F.}~\bibnamefont{Lechermann}},
  \bibinfo{author}{\bibfnamefont{A.}~\bibnamefont{Georges}},
  \bibinfo{author}{\bibfnamefont{F.}~\bibnamefont{Jollet}},
  \bibinfo{author}{\bibfnamefont{T.~O.} \bibnamefont{Wehling}},
  \bibnamefont{and} \bibinfo{author}{\bibfnamefont{A.~I.}
  \bibnamefont{Lichtenstein}}, \bibinfo{journal}{Phys. Rev. B}
  \textbf{\bibinfo{volume}{77}}, \bibinfo{pages}{205112}
  (\bibinfo{year}{2008}).

\bibitem[{\citenamefont{Anisimov et~al.}(2005)\citenamefont{Anisimov, Kondakov,
  Kozhevnikov, Nekrasov, Pchelkina, Allen, Mo, Kim, Metcalf, Suga
  et~al.}}]{ani05}
\bibinfo{author}{\bibfnamefont{V.~I.} \bibnamefont{Anisimov}},
  \bibinfo{author}{\bibfnamefont{D.~E.} \bibnamefont{Kondakov}},
  \bibinfo{author}{\bibfnamefont{A.~V.} \bibnamefont{Kozhevnikov}},
  \bibinfo{author}{\bibfnamefont{I.~A.} \bibnamefont{Nekrasov}},
  \bibinfo{author}{\bibfnamefont{Z.~V.} \bibnamefont{Pchelkina}},
  \bibinfo{author}{\bibfnamefont{J.~W.} \bibnamefont{Allen}},
  \bibinfo{author}{\bibfnamefont{S.-K.} \bibnamefont{Mo}},
  \bibinfo{author}{\bibfnamefont{H.-D.} \bibnamefont{Kim}},
  \bibinfo{author}{\bibfnamefont{P.}~\bibnamefont{Metcalf}},
  \bibinfo{author}{\bibfnamefont{S.}~\bibnamefont{Suga}}, \bibnamefont{et~al.},
  \bibinfo{journal}{Phys. Rev. B} \textbf{\bibinfo{volume}{71}},
  \bibinfo{pages}{125119} (\bibinfo{year}{2005}).

\bibitem[{\citenamefont{Rubtsov et~al.}(2005)\citenamefont{Rubtsov, Savkin, and
  Lichtenstein}}]{rub05}
\bibinfo{author}{\bibfnamefont{A.~N.} \bibnamefont{Rubtsov}},
  \bibinfo{author}{\bibfnamefont{V.~V.} \bibnamefont{Savkin}},
  \bibnamefont{and} \bibinfo{author}{\bibfnamefont{A.~I.}
  \bibnamefont{Lichtenstein}}, \bibinfo{journal}{Phys. Rev. B}
  \textbf{\bibinfo{volume}{72}}, \bibinfo{pages}{035122}
  (\bibinfo{year}{2005}).

\bibitem[{\citenamefont{Werner et~al.}(2006)\citenamefont{Werner, Comanac, de'
  Medici, Troyer, and Millis}}]{wer06}
\bibinfo{author}{\bibfnamefont{P.}~\bibnamefont{Werner}},
  \bibinfo{author}{\bibfnamefont{A.}~\bibnamefont{Comanac}},
  \bibinfo{author}{\bibfnamefont{L.}~\bibnamefont{de' Medici}},
  \bibinfo{author}{\bibfnamefont{M.}~\bibnamefont{Troyer}}, \bibnamefont{and}
  \bibinfo{author}{\bibfnamefont{A.~J.} \bibnamefont{Millis}},
  \bibinfo{journal}{Phys. Rev. Lett.} \textbf{\bibinfo{volume}{97}},
  \bibinfo{pages}{076405} (\bibinfo{year}{2006}).

\bibitem[{\citenamefont{Parcollet et~al.}(2015)\citenamefont{Parcollet,
  Ferrero, Ayral, Hafermann, Krivenko, Messio, and Seth}}]{par15}
\bibinfo{author}{\bibfnamefont{O.}~\bibnamefont{Parcollet}},
  \bibinfo{author}{\bibfnamefont{M.}~\bibnamefont{Ferrero}},
  \bibinfo{author}{\bibfnamefont{T.}~\bibnamefont{Ayral}},
  \bibinfo{author}{\bibfnamefont{H.}~\bibnamefont{Hafermann}},
  \bibinfo{author}{\bibfnamefont{I.}~\bibnamefont{Krivenko}},
  \bibinfo{author}{\bibfnamefont{L.}~\bibnamefont{Messio}}, \bibnamefont{and}
  \bibinfo{author}{\bibfnamefont{P.}~\bibnamefont{Seth}},
  \bibinfo{journal}{Comput. Phys. Commun.} \textbf{\bibinfo{volume}{196}},
  \bibinfo{pages}{398} (\bibinfo{year}{2015}).

\bibitem[{\citenamefont{Seth et~al.}(2016)\citenamefont{Seth, Krivenko,
  Ferrero, and Parcollet}}]{set16}
\bibinfo{author}{\bibfnamefont{P.}~\bibnamefont{Seth}},
  \bibinfo{author}{\bibfnamefont{I.}~\bibnamefont{Krivenko}},
  \bibinfo{author}{\bibfnamefont{M.}~\bibnamefont{Ferrero}}, \bibnamefont{and}
  \bibinfo{author}{\bibfnamefont{O.}~\bibnamefont{Parcollet}},
  \bibinfo{journal}{Comput. Phys. Commun.} \textbf{\bibinfo{volume}{200}},
  \bibinfo{pages}{274} (\bibinfo{year}{2016}).

\bibitem[{\citenamefont{Anisimov et~al.}(1993)\citenamefont{Anisimov, Solovyev,
  Korotin, Czy$\dot{\text{z}}$yk, and Sawatzky}}]{ani93}
\bibinfo{author}{\bibfnamefont{V.~I.} \bibnamefont{Anisimov}},
  \bibinfo{author}{\bibfnamefont{I.~V.} \bibnamefont{Solovyev}},
  \bibinfo{author}{\bibfnamefont{M.~A.} \bibnamefont{Korotin}},
  \bibinfo{author}{\bibfnamefont{M.~T.} \bibnamefont{Czy$\dot{\text{z}}$yk}},
  \bibnamefont{and} \bibinfo{author}{\bibfnamefont{G.~A.}
  \bibnamefont{Sawatzky}}, \bibinfo{journal}{Phys. Rev. B}
  \textbf{\bibinfo{volume}{48}}, \bibinfo{pages}{16929} (\bibinfo{year}{1993}).

\bibitem[{\citenamefont{Sobota et~al.}(2013)\citenamefont{Sobota, Kim, Takatsu,
  Hashimoto, Mo, Hussain, Oguchi, Shishidou, Maeno, Min et~al.}}]{sob13}
\bibinfo{author}{\bibfnamefont{J.~A.} \bibnamefont{Sobota}},
  \bibinfo{author}{\bibfnamefont{K.}~\bibnamefont{Kim}},
  \bibinfo{author}{\bibfnamefont{H.}~\bibnamefont{Takatsu}},
  \bibinfo{author}{\bibfnamefont{M.}~\bibnamefont{Hashimoto}},
  \bibinfo{author}{\bibfnamefont{S.-K.} \bibnamefont{Mo}},
  \bibinfo{author}{\bibfnamefont{Z.}~\bibnamefont{Hussain}},
  \bibinfo{author}{\bibfnamefont{T.}~\bibnamefont{Oguchi}},
  \bibinfo{author}{\bibfnamefont{T.}~\bibnamefont{Shishidou}},
  \bibinfo{author}{\bibfnamefont{Y.}~\bibnamefont{Maeno}},
  \bibinfo{author}{\bibfnamefont{B.~I.} \bibnamefont{Min}},
  \bibnamefont{et~al.}, \bibinfo{journal}{Phys. Rev. B}
  \textbf{\bibinfo{volume}{88}}, \bibinfo{pages}{125109}
  (\bibinfo{year}{2013}).

\bibitem[{\citenamefont{Pavarini et~al.}(2004)\citenamefont{Pavarini, Biermann,
  Poteryaev, Lichtenstein, Georges, and Andersen}}]{pav04}
\bibinfo{author}{\bibfnamefont{E.}~\bibnamefont{Pavarini}},
  \bibinfo{author}{\bibfnamefont{S.}~\bibnamefont{Biermann}},
  \bibinfo{author}{\bibfnamefont{A.}~\bibnamefont{Poteryaev}},
  \bibinfo{author}{\bibfnamefont{A.~I.} \bibnamefont{Lichtenstein}},
  \bibinfo{author}{\bibfnamefont{A.}~\bibnamefont{Georges}}, \bibnamefont{and}
  \bibinfo{author}{\bibfnamefont{O.~K.} \bibnamefont{Andersen}},
  \bibinfo{journal}{Phys. Rev. Lett.} \textbf{\bibinfo{volume}{92}},
  \bibinfo{pages}{176403} (\bibinfo{year}{2004}).

\bibitem[{\citenamefont{Ricc{\`o} et~al.}(2018)\citenamefont{Ricc{\`o}, Kim,
  Tamai, Walker, Bruno, Cucchi, Cappelli, Besnard, Kim, Dudin et~al.}}]{ric18}
\bibinfo{author}{\bibfnamefont{S.}~\bibnamefont{Ricc{\`o}}},
  \bibinfo{author}{\bibfnamefont{M.}~\bibnamefont{Kim}},
  \bibinfo{author}{\bibfnamefont{A.}~\bibnamefont{Tamai}},
  \bibinfo{author}{\bibfnamefont{S.~M.} \bibnamefont{Walker}},
  \bibinfo{author}{\bibfnamefont{F.~Y.} \bibnamefont{Bruno}},
  \bibinfo{author}{\bibfnamefont{I.}~\bibnamefont{Cucchi}},
  \bibinfo{author}{\bibfnamefont{E.}~\bibnamefont{Cappelli}},
  \bibinfo{author}{\bibfnamefont{C.}~\bibnamefont{Besnard}},
  \bibinfo{author}{\bibfnamefont{T.~K.} \bibnamefont{Kim}},
  \bibinfo{author}{\bibfnamefont{P.}~\bibnamefont{Dudin}},
  \bibnamefont{et~al.}, \bibinfo{journal}{Nat. Commun.}
  \textbf{\bibinfo{volume}{9}}, \bibinfo{pages}{4535} (\bibinfo{year}{2018}).

\bibitem[{\citenamefont{Sunko et~al.}(2019)\citenamefont{Sunko, Morales,
  Markovi{\'c}, Barber, Milosavljevi{\'c}, Mazzola, Sokolov, Kikugawa, Cacho,
  Dudin et~al.}}]{sun19}
\bibinfo{author}{\bibfnamefont{V.}~\bibnamefont{Sunko}},
  \bibinfo{author}{\bibfnamefont{E.~A.} \bibnamefont{Morales}},
  \bibinfo{author}{\bibfnamefont{I.}~\bibnamefont{Markovi{\'c}}},
  \bibinfo{author}{\bibfnamefont{M.~E.} \bibnamefont{Barber}},
  \bibinfo{author}{\bibfnamefont{D.}~\bibnamefont{Milosavljevi{\'c}}},
  \bibinfo{author}{\bibfnamefont{F.}~\bibnamefont{Mazzola}},
  \bibinfo{author}{\bibfnamefont{D.~A.} \bibnamefont{Sokolov}},
  \bibinfo{author}{\bibfnamefont{N.}~\bibnamefont{Kikugawa}},
  \bibinfo{author}{\bibfnamefont{C.}~\bibnamefont{Cacho}},
  \bibinfo{author}{\bibfnamefont{P.}~\bibnamefont{Dudin}},
  \bibnamefont{et~al.}, \bibinfo{journal}{arXiv:1903.09581}
  (\bibinfo{year}{2019}).

\bibitem[{\citenamefont{Barber et~al.}(2019)\citenamefont{Barber, Lechermann,
  Streltsov, Skornyakov, Ghosh, Ramshaw, Kikugawa, Sokolov, Mackenzie, Hicks
  et~al.}}]{bar19}
\bibinfo{author}{\bibfnamefont{M.~E.} \bibnamefont{Barber}},
  \bibinfo{author}{\bibfnamefont{F.}~\bibnamefont{Lechermann}},
  \bibinfo{author}{\bibfnamefont{S.~V.} \bibnamefont{Streltsov}},
  \bibinfo{author}{\bibfnamefont{S.~L.} \bibnamefont{Skornyakov}},
  \bibinfo{author}{\bibfnamefont{S.}~\bibnamefont{Ghosh}},
  \bibinfo{author}{\bibfnamefont{B.~J.} \bibnamefont{Ramshaw}},
  \bibinfo{author}{\bibfnamefont{N.}~\bibnamefont{Kikugawa}},
  \bibinfo{author}{\bibfnamefont{D.~A.} \bibnamefont{Sokolov}},
  \bibinfo{author}{\bibfnamefont{A.~P.} \bibnamefont{Mackenzie}},
  \bibinfo{author}{\bibfnamefont{C.~W.} \bibnamefont{Hicks}},
  \bibnamefont{et~al.}, \bibinfo{journal}{arXiv:1909.02743}
  (\bibinfo{year}{2019}).

\bibitem[{\citenamefont{Brahlek et~al.}(2019)\citenamefont{Brahlek, Rimal, Ok,
  Mukherjee, Mazza, Lu, Lee, Ward, Unocic, Eres et~al.}}]{bra19}
\bibinfo{author}{\bibfnamefont{M.}~\bibnamefont{Brahlek}},
  \bibinfo{author}{\bibfnamefont{G.}~\bibnamefont{Rimal}},
  \bibinfo{author}{\bibfnamefont{J.~M.} \bibnamefont{Ok}},
  \bibinfo{author}{\bibfnamefont{D.}~\bibnamefont{Mukherjee}},
  \bibinfo{author}{\bibfnamefont{A.~R.} \bibnamefont{Mazza}},
  \bibinfo{author}{\bibfnamefont{Q.}~\bibnamefont{Lu}},
  \bibinfo{author}{\bibfnamefont{H.~N.} \bibnamefont{Lee}},
  \bibinfo{author}{\bibfnamefont{T.~Z.} \bibnamefont{Ward}},
  \bibinfo{author}{\bibfnamefont{R.~R.} \bibnamefont{Unocic}},
  \bibinfo{author}{\bibfnamefont{G.}~\bibnamefont{Eres}}, \bibnamefont{et~al.},
  \bibinfo{journal}{Phys. Rev. Materials} \textbf{\bibinfo{volume}{3}},
  \bibinfo{pages}{093401} (\bibinfo{year}{2019}).

\bibitem[{\citenamefont{Mazzola et~al.}(2018)\citenamefont{Mazzola, Sunko,
  Khim, Rosner, Kushwaha, Clark, Bawden, Markovi{\'c}, Kim, Hoesch
  et~al.}}]{maz18}
\bibinfo{author}{\bibfnamefont{F.}~\bibnamefont{Mazzola}},
  \bibinfo{author}{\bibfnamefont{V.}~\bibnamefont{Sunko}},
  \bibinfo{author}{\bibfnamefont{S.}~\bibnamefont{Khim}},
  \bibinfo{author}{\bibfnamefont{H.}~\bibnamefont{Rosner}},
  \bibinfo{author}{\bibfnamefont{P.}~\bibnamefont{Kushwaha}},
  \bibinfo{author}{\bibfnamefont{O.~J.} \bibnamefont{Clark}},
  \bibinfo{author}{\bibfnamefont{L.}~\bibnamefont{Bawden}},
  \bibinfo{author}{\bibfnamefont{I.}~\bibnamefont{Markovi{\'c}}},
  \bibinfo{author}{\bibfnamefont{T.~K.} \bibnamefont{Kim}},
  \bibinfo{author}{\bibfnamefont{M.}~\bibnamefont{Hoesch}},
  \bibnamefont{et~al.}, \bibinfo{journal}{PNAS} \textbf{\bibinfo{volume}{115}},
  \bibinfo{pages}{12956} (\bibinfo{year}{2018}).

\bibitem[{\citenamefont{Li et~al.}(2019)\citenamefont{Li, Lee, Wang, Osada,
  Crossley, Lee, Cui, Hikita, and Wang}}]{li19}
\bibinfo{author}{\bibfnamefont{D.}~\bibnamefont{Li}},
  \bibinfo{author}{\bibfnamefont{K.}~\bibnamefont{Lee}},
  \bibinfo{author}{\bibfnamefont{B.~Y.} \bibnamefont{Wang}},
  \bibinfo{author}{\bibfnamefont{M.}~\bibnamefont{Osada}},
  \bibinfo{author}{\bibfnamefont{S.}~\bibnamefont{Crossley}},
  \bibinfo{author}{\bibfnamefont{H.~R.} \bibnamefont{Lee}},
  \bibinfo{author}{\bibfnamefont{Y.}~\bibnamefont{Cui}},
  \bibinfo{author}{\bibfnamefont{Y.}~\bibnamefont{Hikita}}, \bibnamefont{and}
  \bibinfo{author}{\bibfnamefont{H.~Y.} \bibnamefont{Wang}},
  \bibinfo{journal}{Nature} \textbf{\bibinfo{volume}{572}},
  \bibinfo{pages}{624} (\bibinfo{year}{2019}).

\end{thebibliography}

\end{document}